\documentclass[intlimits,twoside,a4paper]{article}

\usepackage{amsmath,amssymb}
\usepackage{graphicx,psfrag}

\usepackage[T2A]{fontenc} 
\usepackage[cp1251]{inputenc} 

\usepackage[eqsecnum]{cmpj2}

\issue{2012}{15}{3}{33603}
\doinumber{10.5488/CMP.15.33603}

\title[Disorder effects on the static scattering function of star branched polymers]%
{Disorder effects on the static scattering function of star branched polymers}
\author[V. Blavatska, C. von Ferber, Yu. Holovatch]{V. Blavatska\refaddr{label1}, C. von Ferber\refaddr{label2,label3}, Yu. Holovatch\refaddr{label1}}
\addresses{
\addr{label1} Institute for Condensed Matter Physics of the National Academy of Sciences of Ukraine,  79011 Lviv, Ukraine
\addr{label2} Applied Mathematics Research Centre, Coventry University, CV1 5FB Coventry, UK
\addr{label3} Institut f\"ur Theoretische Physik II, Heinrich-Heine Universit\"at
D\"usseldorf,
D--40225 D\"usseldorf, Germany
}

\date{Received May 24, 2012, in final form June 26, 2012}

\authorcopyright{V. Blavatska, C. von Ferber, Yu. Holovatch, 2012}

\begin{document}

\maketitle

\begin{abstract}

We present an analysis of the impact of structural disorder on the static scattering function of $f$-armed star branched polymers in $d$ dimensions. To this end, we consider the model of a star polymer immersed in a good solvent  in the presence of structural defects,
correlated at large distances $r$ according to a power law  $ \sim r^{-a}$. In particular, we are interested in the ratio $g(f)$
of the radii of gyration of star and linear polymers of the same molecular weight, which is a universal experimentally measurable quantity.
We apply a direct polymer renormalization approach and evaluate the results within the double $\varepsilon=4-d$, $\delta=4-a$-expansion.
We find an increase of $g(f)$ with an increasing $\delta$. Therefore, an increase of disorder correlations leads to an increase of the size measure of a star relative to linear polymers of the same molecular weight.
\keywords  polymers, structural disorder, universality, renormalization group
\pacs 61.41.+e, 61.25.hp, 64.60.ae
\end{abstract}

\vspace{2ex}
\section{Introduction}

Scattering experiments have been commonly used in investigations of the structure properties of condensed
matter for more than a century (see, e.g.~\cite{desCloizeaux}). For polymer systems, the quantity of interest is the static structure function
$S({{k}})$ as a function of the wave vector ${\vec{k}}$,  representing the Fourier transform of the monomer-monomer
correlation function~\cite{Roovers72,Roovers83,Huber84,Khasat88,Bauer89,Merkle93}.
The scattering intensity $I(k)\equiv S(k)/S(0)$ at small  $k$
gives the radius of gyration $R_{g}$ of a single macromolecule:
\begin{equation}
   I(k)= 1-k^2 \frac{\langle R_{g}^2 \rangle}{d}+\ldots, \label{int}
\end{equation}
where $d$ is the space dimension and $\langle \ldots \rangle$ denotes the average over the ensemble of all conformations that a macromolecule in a solvent is capable of attaining.

\begin{figure}[ht]
\begin{center}
\includegraphics[width=4cm]{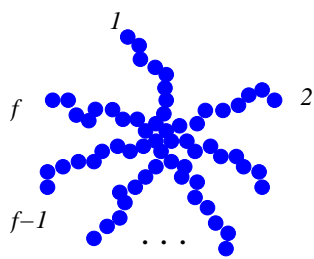}
\caption{Schematic presentation of an $f$-armed star polymer.}
\label{perc}
\end{center}
\end{figure}

In this paper, we intend to  derive some theoretical predictions
and quantitatively describe the peculiarities of scattering experiments with star-like polymers.
Star-like polymers are the simplest representatives of a class of branched polymer structures
that are in a close relationship to complex systems such as gel, rubber, micellar and
other polymeric and surfactant systems~\cite{Grest96,Likos01,Ferber02}. In particular, some conformational properties
of star polymers could be easily generalized to determine the behavior of
polymer networks of a more complicated structure~\cite{Duplantier89,schafer92}.
The star polymer can be viewed as $f$ linear polymer chains (arms) linked together at the central core (see figure~\ref{perc}).
For $f=1(2)$, one restores the polymer chain of linear architecture, whereas it has been shown that in another limiting situation ($f\gg 1$), the star polymer
attains the features of a soft colloidal particle~\cite{Lowen,Lowen_2}.

A convenient parameter for comparing the size measure of a star consisting of $f$ arms (each of length $N$) and a linear polymer chain
having the same molecular weight $fN$
is the ratio:
 \begin{equation}
 g(f)\equiv \frac { \langle R^2_{g{\rm star}} \rangle}{  \langle R^2_{g{\rm chain}} \rangle}\,, \label{gratio}
 \end{equation}
 here, $\langle R^2_{g{\rm star}} \rangle$ is the mean square radius of gyration of the star polymer.
 Note, that the value~(\ref{gratio}) is an experimentally measurable quantity. Indeed, recalling~(\ref{int}), one has at small $k$:
 \begin{equation}
 \frac{I'(k)_{{\rm star}}}{I'(k)_{{\rm chain}}}=g(f)+\ldots \,.\label{exper}
 \end{equation}
 Therefore,  the ratio of the derivatives $I'(k)$ of the corresponding scattering intensities for small values of the wave vector permits to  experimentally define the $g(f)$ ratio.
Moreover, it is well established that
the gyration radius of a star polymer scales with its total number of segments
according to the scaling law:
\begin{equation}
\langle R^2_{g{\rm star}} \rangle \sim (fN)^{2\nu},\label{nu1}
\end{equation}
with an universal exponent $\nu$ that also governs the scaling of a single polymer chain of $N$ monomers: $\langle R^2_{g{\rm chain}} \rangle\sim N^{2\nu}$.
Therefore, the ratio $g(f)$ of a star and linear chain of the same total molecular weight is $N$-independent.

In the pioneering work by Zimm and Stockmayer~\cite{Zimm49}, an estimate for the size ratio $g(f)$ was found analytically:
\begin{equation}
g(f)=\frac{3f-2}{f^2}\,. \label{zimm}
\end{equation}
Inserting $f=1$ or $f=2$ in this relation,  one restores the trivial result $g=1$. For any $f\geqslant 3$, ratio  (\ref{zimm})
is smaller than $1$, reflecting the fact that the size of a branched polymer is always smaller than the size of a linear polymer chain of the same molecular weight.
Note that in deriving the expression~(\ref{zimm}) the excluded volume effect was neglected (restricting to the idealized Gaussian case, where~(\ref{nu1}) holds with size exponent $\nu=1/2$).
In this limit, the result holds
for any space dimension. Dimensional dependence of the size ratio $g(f)$ is found by introducing the concept of
excluded volume applied  to polymer macromolecules by P.~Flory. It refers to the
idea that any segment (monomer) of macromolecule is not capable of occupying the space that is already occupied by another segment;
this causes a swelling of a polymer in  a solution with size exponent $\nu(d)\geqslant 1/2$.
Later on, analytical~\cite{Daoud82,Miyake82,Miyake82_2,Alessandrini92} and numerical~\cite{Whittington86,Grest87,Batouslis22,Bishop93,Wei97}
studies have found the value of
$g(f)$ to increase if the excluded volume effect is taken into consideration.

While reliable estimates are known for the value of $g(f)$ for polymers in a good solvent, there are no similar
 estimates for the case when the polymer is immersed in a good solvent in the presence of structural impurities or a porous environment.
 However,  such estimates are of a great importance
for understanding the behavior of macromolecules in colloidal solutions~\cite{Pusey86} or near microporous membranes~\cite{Cannel80}. The density fluctuations of such obstacles  lead to large spatial
inhomogeneities which often produce pore-like fractal structures~\cite{Dullen79}.
Such a disordered (porous) environment may be found, in particular, in a biological cell
composed of many different kinds of biochemical species~\cite{Goodesel91,Goodesel91_2,Goodesel91_3}.
It has been proven both analytically~\cite{Kim} and numerically~\cite{Lee,Lee_2} that the
presence of  uncorrelated point-like defects of weak concentration does not change the universality class of polymers.
Here, however, we address a case where the structural obstacles of the environment are spatially
correlated on a mesoscopic scale~\cite{Sahimi95}.
Following reference~\cite{Weinrib83}, this case may be described by assuming the defects to be correlated at large distances $r$
according to a power law resulting in a pair correlation function:
\begin{equation}
h(r)\sim r^{-a}. \label{func}
\end{equation}
For $a<d$, specific situations that give rise to such a correlation function include the defects extended in space, e.g., the cases $a=d-1$ ($a=d-2$)  may be represented by lines (planes) of defects of random orientation, whereas non-integer values of $a$ may include obstacles of fractal structures (see~\cite{Weinrib83,Blavatska01a,Blavatska01a_2} for further details).
The impact of long-range-correlated disorder on the scaling of linear and star branched polymers has been analyzed in
 previous works~\cite{Blavatska01a,Blavatska01a_2} by means of the
 field-theoretical renormalization group (RG) approach.
The aim of the present paper is to  analytically evaluate an experimentally measurable ratio~(\ref{exper}) for a star polymer in a solvent in the presence of structural defects correlated according to~(\ref{func}).

The layout of the reminder of the paper is as follows. In the next section, we develop a description of the problem in the frames of the Edwards
continuous  chain model. In section~3, a  direct polymer renormalization method is briefly described; the results of its application to the
problem under consideration are presented in section~4.
Conclusions and an outlook are given in section~5.

\section{The model}

We consider  star polymers with $f$ arms in a solution in the presence of structural obstacles.  In the frames of
the Edwards continuous chain model~\cite{desCloizeaux}, each arm of the star is presented by a path $ r_i(s)$,
parameterized by $0\leqslant s\leqslant S_i$, $i=1,2,\ldots,f$. The central branching point of the star is fixed in space,
so that: $ \vec{r}_1(0)=\ldots=\vec{r}_f(0)=0$. We take the contour length of all arms to be equal: $S_1=\ldots =S_f=S$.
The partition function of the system reads:
\begin{eqnarray}
{\cal Z}_f(S)&=&\int {\cal D} \{r\}\,
\prod_{i=1}^f\delta\left( \vec{r}_i(0)\right) \exp\left \{-\frac{1}{2}\sum_{i=1}^f\int_0^{S}{\rm d}s
\left[\frac{{\rm d} {\vec  r}_{i}(s)}{{\rm d} s}\right]^2\right.\nonumber\\
&&{}-\left.\frac{b_0}{2}\sum_{i,j=1}^f
\int_0^{S}{\rm d}s'\int_0^{S}{\rm d}\,s{''}\,\delta\left[\vec{r}_i(s'')-\vec{r}_j(s{'})\right]+\sum_{i=1}^f \int_0^{S}{\rm d}s\,V[\vec{r}_i(s)]\right\}.\label{pure}
\end{eqnarray}
Here, a multiple path integral is performed for the paths $r_1,\ldots,r_f$ and
the product of $\delta$-functions reflects the star-like configuration of $f$ chains.
The first term in the exponent represents the chain connectivity, the
second term describes the short range excluded volume interaction with a bare coupling constant $b_0$,
and the last term contains a random potential $V[\vec{r}_i(s)]$  arising due to the presence of structural disorder.
Denoting by ${\overline{(\ldots)}}$ the average over different realizations of disorder,  the first moment of the distribution  is:
$${\overline { V[{\vec {r}}(s)]} } =\rho_0$$ with $\rho_0$ being the density of obstacles.
 Let us introduce a notation for the second moment: \begin{equation} {\overline{ V[\vec{r}(s)]V[\vec{r}(s')]}} \equiv h\left[\vec{r}(s)-\vec{r}(s')\right].\label{avv0}
\end{equation}

Note that  dealing with systems that display randomness of structure, one usually encounters two types of ensemble averaging treated as quenched and annealed disorder~\cite{Brout59,Brout59_2}. The annealed case amounts to averaging  the partition sum of a system over the random variables, whereas in the
quenched case, the free energy (or the logarithm of the partition sum) is to be averaged; the replica formalism is usually applied in the last situation.
In principle, the behavior of systems with quenched and annealed disorder is quite different. However, as it has been shown in a number of works~\cite{Cherayil90,Wu91,Ippolito98,Patel03},
{\it the distinction between quenched and annealed averages for an infinitely long
single polymer chain is negligible},
and in performing analytical calculations for quenched polymer systems one may thus restrict the problem  to the simpler case of annealed averaging.
To average the partition function of a system over different realizations of
obstacles, we make use of the relation:
\begin{equation}
\overline {{\rm e}^{ax}}=\exp \left[ \sum\limits_{n=1}^{\infty}\frac{a^nM_n(x)}{n!}  \right],
\end{equation}
where $M_n(x)$ are $n$th cumulants of the random variable $x$: $M_1(x)={\overline{ x}}$, $M_2(x)={\overline{(x-\overline{x})^2}}$ etc.
Noticing that only the last term in (\ref{pure}) contains random variables and taking into account (\ref{avv0}), we obtain:
\begin{eqnarray}
&&{\overline {{\cal Z}_f(S)}} =
\int {\cal D} \{r\}\,
\prod_{i=1}^f\delta[ \vec{r}_i(0)] {\rm e}^{-H_{{\rm dis}}} \label{Zdis}
\end{eqnarray}
with an effective Hamiltonian:
\begin{eqnarray}
H_{{\rm dis}}&=&\frac{1}{2}\sum_{i=1}^f\int_0^{S}{\rm d}s
\left[\frac{{\rm d}  {\vec {r}}_{i}(s)}{{\rm d} s}\right]^2+\frac{b_0}{2}\sum_{i,j=1}^f
\int_0^{S}{\rm d}s'\int_0^{S}{\rm d}\,s{''}\,\delta\left[{{\vec{r}}_i}(s'')-{\vec{r}}_j(s{'})\right]\nonumber\\
&&{}- \frac{1}{2}\sum_{i=1}^f \int_0^{S}{\rm d}s'\int_0^{S}{\rm d}\,s{''}\,h\left[{{\vec{r}}_i}(s'')-{\vec{r}}_j(s{'})\right]-\rho_0fS-\frac{1}{2}\rho_0^2fS^2.
\label{Hdis}
\end{eqnarray}
The last two terms in (\ref{Hdis}) correspond to a trivial constant shift which will be omitted in the following analysis.
Note also, that in (\ref{Hdis}) we do not take into account the terms generated by higher-order correlations of the type (\ref{avv0}),
because for the problem under consideration these terms are irrelevant in the renormalization group sense.

The case of structural disorder in the form of point-like uncorrelated defects corresponds to $h[{{\vec{r}}}(s'')-{\vec{r}}(s{'})]=v_0\delta[{{\vec{r}}}(s'')-{\vec{r}}(s{'})] $  where $v_0$ is some constant.
One immediately reveals that in this case one can adsorb the effect of disorder into the excluded volume coupling constant
 passing to the coupling: $b_0\equiv b_0-v_0$. This conclusion was obtained for the case of polymers in quenched disorder
 by Kim~\cite{Kim} based on  a refined field-theoretical study; in the present case of annealed disorder, this is a  straightforward result.

  We address the model where the
structural obstacles  are spatially
correlated  at large distances $r$ according to (\ref{func}).
Taking into account that the Fourier transform of the correlation function at small $k$ is related to its large-$r$ behaviour via:
\begin{equation} \label{fff}
h\left[\left|\vec{r}_i(s'')-\vec{r}_j(s')\right|\right]\cong \left|\vec{r}_i(s'')-\vec{r}_j(s')\right|^{-a}\cong w_o\int{\rm d}k\,k^{a-d}\,{\rm e}^{\ri\vec{k}\left[\vec{r}_i(s'')-
\vec{r}_j(s')\right]},
\end{equation}
one is left with a model with two couplings $b_0$ and $w_0$.
Note that coupling $b_0$ should be positive, which corresponds to an effective mutual repulsion of the monomers due to the excluded volume effect. The coupling $w_0$ is positive
as results from the Fourier image of the correlation function.

Performing dimensional analysis for the  terms in~(\ref{Hdis}), one finds the dimensions of the couplings in terms of a dimension of contour length $S$: $[b_0]=[S]^{{\rm d}_{b_0}}$, $[w_0]=[S]^{{\rm d}_{w_0}}$
with ${\rm d}_{b_0}=(4-d)/2$, ${\rm d}_{w_0}=(4-a)/2$. Note that for the exponent in (2.1) to be dimensionless, the contour length
 needs to have units of surface. The ``upper critical'' values of the space dimension ($d_{\mathrm{c}}=4$) and the correlation parameter ($a_{\mathrm{c}}=4$), at which the couplings are dimensionless, play an important role in the renormalization scheme, as outlined below.

\section{The method}

To study the universal properties of polymer macromolecules in solutions, it is convenient to apply the direct
renormalization method, as developed by des Cloizeaux~\cite{desCloizeaux}. The efficiency of this approach comes, on the one hand, from
its close relation to the concepts of field theory~\cite{rg,rg_2,rg_3},
and, on another hand, from providing a considerably simpler treatment of a variety of complex polymer systems.

In the asymptotic limit of an infinite linear measure of a continuous polymer curve (corresponding to an infinite  number of
  configurations), one observes various divergences. All these divergences can be eliminated by
introducing corresponding renormalization factors directly associated with physical quantities. This postulates the existence of a limiting theory
that describes the sets of very long polymers.

As a first step within this theory, we consider the size measure of an $f$-arm star polymer given by
  the mean square end-to-end distance of its individual arm. When evaluated in terms of a perturbation theory series in
bare coupling constants $\{\lambda_0\}$, this reads:
\begin{equation}
\langle R_{\mathrm{e}}^2 \rangle =\langle [\vec{r}(S)-\vec{r}(0)]^2\rangle=\chi_0(\{ \lambda_0\})S.\label{chi0def}
\end{equation}
Here, the averaging is performed with respect to a corresponding effective Hamiltonian, and $\chi_0(\{ \lambda_0\})$ is the so-called swelling factor that reflects
the impact of interactions on the effective size of macromolecules. For the case of a Gaussian chain (all couplings $\lambda_0=0$), one has $\chi_0(\{0\})=1$.
Recalling the scaling of a polymer size with its molecular weight:
\begin{equation}
\langle R_{\mathrm{e}}^2 \rangle \sim N^{2\nu} \sim S^{2\nu},
\end{equation}
one finds an estimate for the effective critical exponent $\nu(\{\lambda_0 \})$:
\begin{equation}
2\nu(\{\lambda_0 \})-1=S\frac{\partial }{\partial S}\ln \chi_0(\{\lambda_0 \}). \label{nuexp}
\end{equation}

The second renormalization factor $\chi_1(\{\lambda_0 \})$ is introduced via:
\begin{equation}
\frac{{\cal Z}_{f}(S)}{{\cal Z}_f^0(S)}=\left[\chi_1(\{\lambda_0 \})\right]^{2}.
\end{equation}
Here, ${\cal Z}_{f}(S)$ is the partition function of an $f$-arm star polymer and
${\cal Z}_f^0(S)$ is the partition function of an idealized Gaussian model.
It is established that the number of all possible conformations of an $f$-armed star polymer scales with the weight of a  macromolecule
parametrised by $S$ as:
\begin{equation}
{\cal Z}_f(S)\sim \mu^{fS}(fS)^{\gamma_f-1}. \label{gamma1}
\end{equation}
Here, the $\gamma_f$ are additional universal critical exponents depending only on the space dimension $d$ and the number of arms $f$
(exponents $\gamma_1=\gamma_2\equiv\gamma$
restore the value for the single polymer chain), $\mu$ is  a non-universal fugacity.
In a similar way  as for the size measure, from the scaling assumption (\ref{gamma1}) one finds
an estimate for an effective critical exponent $\gamma_f(\{\lambda_0 \})$ governing the scaling behavior of the
  number of possible configurations as:
\begin{equation}
\frac{\gamma_f(\{\lambda_0 \})-1}{2}=S\frac{\partial \ln \chi_1(\{\lambda_0 \}) }{\partial S}\,.\label{gammaexp}
\end{equation}
The critical exponents (\ref{nuexp}) and (\ref{gammaexp})  presented in the form of series expansions in the coupling constants
$\{\lambda_0 \}$ are, however, divergent in the asymptotic limit of large $S$.
To eliminate these divergences, renormalization of the coupling constants is performed.
Subsequently, the critical exponents attain finite values when evaluated at a stable fixed point (FP) of the renormalization group transformation.
Note that the FP coordinates are universal. In particular, the scaling of a single polymer chain and that of  a polymer star is governed by the same
unique FP. Therefore, to evaluate the FP coordinates in the following analysis,
we restrict ourselves to a simpler case of a single chain polymers ($f=1$). To define
the coupling constant renormalization,
one considers the second virial coefficient of a polymer solution given by the relation:
\begin{equation}
\Pi\beta=C-\frac{1}{2}C^2\sum_{\lambda_0}\frac{{\cal Z}_{\lambda_0}(S,S)}{[{\cal Z}_{1}(S)]^2}+\ldots,
\end{equation}
here, $\Pi$ is the osmotic pressure, $\beta=1/k_{\mathrm{B}}T$, $C$ is the number of monomers per unit volume, and ${\cal Z}_{1}(S)$ is the
 partition function of a single polymer chain.
${\cal Z}_{\lambda_0}(S,S)$ are contributions into a partition function of two interacting chain polymers having dimensions
 ${\cal Z}_{\lambda_0}(S,S)\sim [S]^2[\lambda_0]$.
The renormalized coupling constants $\lambda_{\mathrm{R}}$ are thus defined by:
\begin{eqnarray}
&&\lambda_{\mathrm{R}}(\{ \lambda_0\})=-\left[\chi_1(\{\lambda_0 \})\right]^{-4}{Z}_{\lambda_0}(L,L)\left[2\pi\chi_0(\{\lambda_0 \})L \right]^{-(2-{\rm d_{\lambda_0}}) }, \label{ures}
\end{eqnarray}
therefore:
\begin{equation}
\Pi\beta=C+\frac{1}{2}\sum_{\lambda_{\mathrm{R}}}\lambda_{\mathrm{R}} C^2\left[2\pi\chi_0(\{\lambda_0 \})L\right]^{(2-{\rm d_{\lambda_0}})}+\ldots \ .
\end{equation}
In the limit of infinite linear size of macromolecules,  the renormalized theory remains finite, such that:
\begin{equation}
\lim_{S\to\infty} \lambda_{\mathrm{R}}(\{ \lambda_0\})=\lambda_{\mathrm{R}}^*\, .
\end{equation}
Moreover, for negative  ${\rm d_{\lambda_0}}\leqslant 0$, macromolecules are expected to behave like Gaussian chains in spite of
the interactions between monomers, thus each $\lambda_{\mathrm{R}}^*=0$ for corresponding ${\rm d_{\lambda_0}}\leqslant 0$.
It is, therefore, proper to choose $\{\lambda_{\mathrm{R}}\}$ as expansion parameters which remain finite for $S\to\infty$ and which are
also rather small close to the critical dimensions of the corresponding couplings. The concept of expansion  in small
deviations from the upper critical dimensions of the coupling constants thus naturally arises.

 The flows of the renormalized coupling constants are governed by functions $\beta_{\lambda_{\mathrm{R}}}$:
\begin{equation}
\beta_{\lambda_{\mathrm{R}}}=2S\frac{\partial \lambda_{\mathrm{R}}(\{ \lambda_0\})}{\partial S}\,.
\end{equation}
 Reexpressing $\{ \lambda_0\}$ in terms of renormalized couplings $\lambda_{\mathrm{R}}$ according to
 (\ref{ures}),  the fixed points of renormalization group transformations are given
 by common zeros of the $\beta$-functions.
 Stable fixed points govern the asymptotical scaling properties of macromolecules in solutions and
 make it possible, e.g., to obtain reliable
 asymptotical values of the critical exponents  (\ref{nuexp}) and (\ref{gammaexp}).

\section{Results}

\begin{figure}[!b]
\begin{center}
\psfrag{sp}{\Large{$s^{'}$}}
\psfrag{spp}{\Large{$s^{''}$}}
\includegraphics[width=10cm]{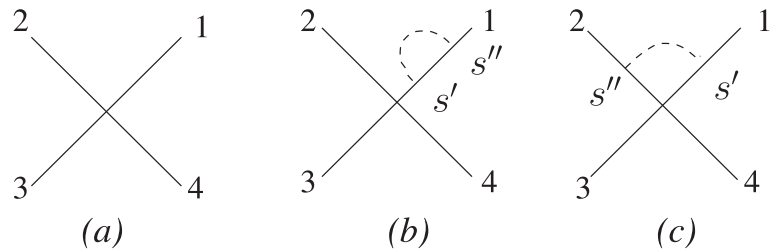}
\end{center}
 \caption{Diagram contributions to the partition function of a $4$-arm star up to the first order in the coupling constants. Dotted lines denote
 possible interactions between points $s'$, $s''$, governed by couplings $b_0$ and $w_0$. Integrations are to be performed over all
 positions of  the segment end points, i.e., over all mutual interaction points within a single arm and between different arms.}
 \label{fig2}
 \end{figure}

 We start by evaluating the partition function (\ref{Zdis}) of the model with an effective Hamiltonian (\ref{Hdis}),
performing an expansion in coupling constants $b_0$, $w_0$:
  \begin{eqnarray}
\overline {Z_f(S)}&=&\int{\cal D} r\,
\exp\left\{ {- \frac{1}{2}\sum\limits_{i=1}^f\int\limits_0^{S}{\rm d}s \left[\frac{{\rm d} r_{i}(s)}{{\rm d} s}\right]^2}\right\}
\left\{1-\frac{b_0}{2} \sum_{i,j=1}^f
\int_0^{S}{\rm d}s'\!\!\int_0^{S}{\rm d}s{''}\!\!\int{\rm d} k\, {\rm e}^{\ri\vec{k}\left[\vec{r}_i(s'')-\vec{r}_j(s{'})\right]}\right.\nonumber\\
&&{}+\left.  \frac{w_0}{2}\sum_{i,j=1}^f
\int_0^{S}{\rm d}s'\int_0^{S}{\rm d}\,s{''}\,\int{\rm d} k\,k^{a-d}\, {\rm e}^{\ri\vec{k}\left[\vec{r} _i(s'')-\vec{r} _j(s{'})\right]}
+\ldots\right\},
\end{eqnarray}
here, the Fourier-transform of the $\delta$-function is exploited and the last term originates from the Fourier transform of the function
$h$ at small $k$  [see equation~(\ref{fff})]. Below, we will consider the one-loop approximation,
keeping only the first-order terms in $b_0$, $w_0$ in the expansions.
One may rewrite:
\begin{eqnarray}
&&{\rm e}^{\ri\vec{k}\left[\vec{r}_i(s'')-\vec{r}_i(s{'})\right]}=
\exp\left[\ri\vec{k}\int_{s'}^{s''}\frac{\rd \vec{r}_i(s)}{\rd\,s}{\rm d}s\right],\\
&&{\rm e}^{\ri\vec{k}\left[\vec{r}_i(s'')-\vec{r}_j(s{'})\right]}=\exp\left[\ri\vec{k}\int_{0}^{s''}\frac{\rd \vec{r}_i(s)}{\rd s}{\rm d}s-
\ri\vec{k}\int_{0}^{s'}\frac{\rd \vec{r}_j(s)}{\rd s}{\rm d}s\right],\label{t1}
\end{eqnarray}
taking into account that $r_i(0)=r_j(0)=0$ in our model of a star-shaped polymer. Making use of the identity:
\begin{eqnarray}
 \exp \left\{-\frac{1}{2}\int_{s'}^{s''}{\rm d}s
\left[\frac{{\rm d} r_{i}(s)}{{\rm d} s}\right]^2 +
\ri\vec{k}\int_{s'}^{s''}\frac{\rd \vec{r}_i(s)}{\rd\,s}{\rm d}s
 \right\}=\exp \left\{-\frac{1}{2}\int_{s'}^{s''}{\rm d}s\left\{
\left[\frac{{\rm d} r_{i}(s)}{{\rm d} s}-\ri\vec{k}\right]^2 +
k^2 \right\}\right\}
 \end{eqnarray}
 and taking into account that:
 \begin{equation}
\int_{-\infty}^{\infty}\rd x {\rm e}^{-A(x-\ri k)^2}=\int_{-\infty}^{\infty}\rd x {\rm e}^{-Ax^2},
\label{t4}
\end{equation}
we receive:
 \begin{eqnarray}
\overline{ {\cal Z}_f(S)}&=& {\cal Z}_f^0(S) \left\{1-b_0(2\pi)^{-\frac{d}{2}}\left[f
\int\limits_0^{S}{\rm d}s''\int\limits_0^{s''}{\rm d}\,s{'}\,\left(s''-s'\right)^{-d/2}\right.\right. \nonumber\\
&&{}+\left.\frac{f(f-1)}{2}
\int\limits_0^{S}{\rm d}s''\int\limits_0^{S}{\rm d}\,s{'}\,
\left(s'+s''\right)^{-d/2} \right]  \nonumber\\
&&{}+\left. w_0(2\pi)^{-\frac{a}{2}}\left[f
\int\limits_0^{S}{\rm d}s''\int\limits_0^{s''}{\rm d}\,s{'}\,\left(s''-s'\right)^{-a/2}\right.\right.
+\left.\left.\frac{f(f-1)}{2}
\int\limits_0^{S}{\rm d}s''\int\limits_0^{S}{\rm d}\,s{'}\,
\left(s'+s''\right)^{-a/2} \right] \right\}.
\label{before}
\end{eqnarray}
In the last equation, the Gaussian integration over $k$ is performed and the notation ${\cal Z}_f^0(S)$ is introduced for the partition function of the ``unperturbed'' Gaussian model:
\begin{equation}
{\cal Z}_f^0(S)=\int{\cal D} r\,
\exp\left\{ {- \frac{1}{2}\sum\limits_{i=1}^f\int\limits_0^{S}{\rm d}s
\left[\frac{{\rm d} r_{i}(s)}{{\rm d} s}\right]^2}\right\}. \label{zz00}
\end{equation}
In what follows, we will use the diagrammatic representation of the perturbation theory series (see figure~\ref{fig2}).
Performing the integrals in (\ref{before}) and introducing dimensionless couplings
\begin{equation}
b=b_0(2\pi)^{-d/2}S^{2-d/2}, \qquad w=w_0(2\pi)^{-a/2}S^{2-a/2}
\end{equation}
we obtain:
\begin{eqnarray}
\overline{{\cal Z}_f(S)}&=&{\cal Z}_f^0(S) \left\{1-\frac{4b}{(2-d)(4-d)}\left[f+ \frac{f(f-1)}{2}(2^{2-d/2}-2)\right]\right.\nonumber\\
&&{}+\left.\frac{4w}{(2-a)(4-a)}\left[f+ \frac{f(f-1)}{2}(2^{2-a/2}-2)\right]\right\}.
\label{Zf}
\end{eqnarray}
Finally, one may perform a double $\varepsilon=4-d$, $\delta=4-a$-expansions:
\begin{eqnarray}
\overline{{\cal Z}_f(S)}&=&1+b\left[\frac{f(3-f)}{\varepsilon}+\frac{f(3-f)}{2}+\frac{f(f-1)}{2}\ln(2) \right]\nonumber\\
&&{}-w\left[\frac{f(3-f)}{\delta}+\frac{f(3-f)}{2}+\frac{f(f-1)}{2}\ln(2) \right].
\label{Zfeps}
\end{eqnarray}

 The averaged squared end-to-end distance $\overline{\langle R^2_{\mathrm{e}} \rangle}$  of a single arm of a star polymer may be calculated using the identity:
\begin{eqnarray}
\label{ras}
\overline{\langle R^2_{\mathrm{e}} \rangle}= \overline{\langle[\vec{r}(S)-\vec{r}(0)]^2\rangle}  =-2d\frac{\partial}{\partial q^2}\overline{\langle{\rm e}^{\ri\vec{q}\,\left[\vec{r}(S)-\vec{r}(0)\right]}\rangle}\big|_{q=0}\,,
\end{eqnarray}
where:
\begin{equation}
\overline{{\langle \ldots \rangle}}=
\frac{\int {\cal D}r\, {\rm e}^{-{\cal H_{{\rm dis}}}} (\ldots)\prod_{i=1}^f\delta[ \vec{r}_i(0)]}{{\overline {{\cal Z}_f(S)}}}\, .
 \label{aver}
\end{equation}
Following the same scheme as described above for the partition function, we find:
\begin{eqnarray}
\overline{\langle R_{\mathrm{e}}^2\rangle } =Sd\left[1+\frac{4b}{\left(4-d\right)\left(6-d\right)}-
\frac{4w}{\left(4-a\right)\left(6-a\right)} \right]. \label{rr}
\end{eqnarray}
We may, therefore, define a swelling factor $\chi_0(b_0,w_0)$ [cf. (\ref{chi0def})] as:
\begin{equation}
\chi_0(b_0,w_0)=\left[1+\frac{4b}{\left(4-d\right)\left(6-d\right)}-
\frac{4w}{\left(4-a\right)\left(6-a\right)} \right]=\left[1+\frac{b}{\varepsilon}(2-\varepsilon)-\frac{w}{\delta}(2-\delta)\right].
\label{Lr}
\end{equation}

\begin{figure}[ht]
\begin{center}
\psfrag{s}{{\Large $s_1$}}
\psfrag{ss}{{\Large $s_2$}}
\includegraphics[width=6cm]{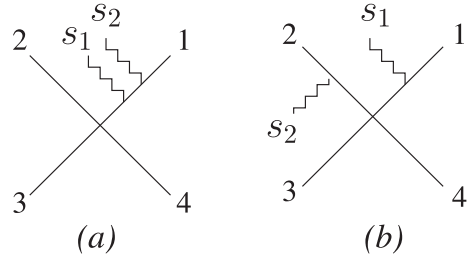}
 \end{center}
 \caption{Diagrammatic presentation of contributions into radius of gyration of star polymer in zeros order of perturbation theory.}
 \label{Rnol}
 \end{figure}
Now we return to the calculation of  the gyration radius.
The gyration radius of a star polymer in a solvent in the presence of correlated defects  is defined by:
\begin{equation}
 \overline{\langle R^2_{g{\rm star}}\rangle }=\frac{1}{2(fS)^2}\int_0^{S}{\rm d} s_1\int_{0}^{S} {\rm d} s_2\, \overline{\left \langle \sum_{i,j=1}^f \left[ \vec{r}_i(s_2)-\vec{r}_j(s_1)\right]^2\right\rangle}. \label{rgdefin}
\end{equation}
We rewrite:
\begin{equation}
\overline{\left\langle \sum_{i,j=1}^f\left[ \vec{r}_i(s_2)-\vec{r}_j(s_1)\right]^2\right\rangle}
=-2{d}\frac{{\partial}}{{\partial} |q|^2}\overline{\left\langle {\rm e}^{\ri\vec{q}\sum_{i,j=1}^f[\vec{r}_i(s_2)-\vec{r}_j(s_1)]}\right\rangle}\bigg|_{q=0}\,,\label{def}
 \end{equation}
 where ${d}$ is the space dimension.
First, let us consider the zero-loop order of the expansion of (\ref{def}) in coupling constants. A diagrammatic representation is given in figure~\ref{Rnol}.
The analytic expression, corresponding to the diagram (a) reads:
\begin{equation}
f\overline{\left\langle {\rm
e}^{\ri\vec{q}[\vec{r}_1(s_2)-\vec{r}_1(s_1)]}\right\rangle}=f {\rm e}^{-\frac{q^2}{2}(s_2-s_1)},
\end{equation}
whereas the diagram (b) gives:
 \begin{equation}
\frac{f(f-1)}{2}\overline{\left\langle {\rm
e}^{\ri\vec{q}[\vec{r}_2(s_2)-\vec{r}_1(s_1)]}\right\rangle}=\frac{f(f-1)}{2} {\rm e}^{-\frac{q^2}{2}(s_1+s_2)}.
 \end{equation}
 Note, that here and in what follows the Gaussian integral in (\ref{zz00}) cancels against the numerator
when averaging~(\ref{aver}) is performed.
One thus distinguishes between two types of contributions: one resulting from insertions
$s_1$, $s_2$ along the same arm of the star, and the second corresponding to insertions located on two different arms. Taking the derivative
with respect to $q$, and evaluating for $q=0$ according to~(\ref{def}), we have for the radius of gyration of a star polymer in the unperturbed (Gaussian) case:
\begin{equation}
 \overline{\langle R^2_{g{\rm star}}\rangle } =\frac{d}{(fS)^2}\left[f\int\limits_0^{S}{\rm d}s_2\int\limits_0^{s_2}{\rm d}s_1\,(s_2-s_1)+
\frac{f(f-1)}{2} \int\limits_0^{S}{\rm d}s_1\int\limits_0^{S}{\rm d}s_2\,(s_1+s_2)\right]=
\frac{dS}{f}\frac{3f-2}{6}\,.
\end{equation}

\begin{figure}[h]
\centerline{
\includegraphics[width=0.98\textwidth]{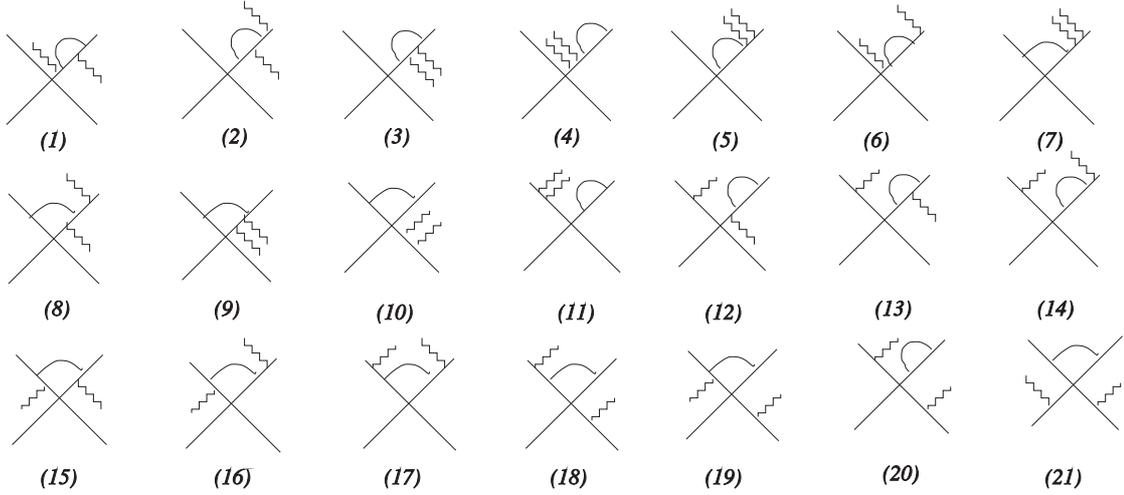}
}
 \caption{ Diagram contribution into the gyration radius at the  one-loop level.}
\label{Rstar}
\end{figure}
Now, let us perform calculations up to the first order of the perturbation theory expansion in the couplings $b_0$, $w_0$.
In figure~\ref{Rstar}, we present diagram contributions to (\ref{def}), see appendix for details.
We have:
\begin{eqnarray}
&&\hspace{-0.5cm}\overline{\left\langle \sum_{i,j=1}^f\left[ \vec{r}_i(s_2)-\vec{r}_j(s_1)\right]^2\right\rangle}
=-2d\frac{\partial}{\partial q^2}\Big\{
f\Big[I_1(d)+I_1(a)+I_2(d)+I_2(a)+I_3(d)\nonumber\\
&&\hspace{1.35cm}+I_3(a)+ I_4(d)+I_4(a)+I_5(d)+I_5(a)+I_6(d)+I_6(a)+I_7(d)+I_7(a)\Big]\nonumber\\
&&\hspace{1.35cm}+\frac{f(f-1)}{2}\Big[I_{8}(d)+I_{8}(a)+I_{9}(d)+I_{9}(a)+I_{10}(d)+I_{10}(a)
+I_{11}(d)+I_{11}(a)\nonumber\\
&&\hspace{1.35cm}+I_{12}(d)+I_{12}(a)+I_{13}(d)+I_{13}(a)+I_{14}(d)+I_{14}(a)+I_{15}(d)+I_{15}(a)
\nonumber\\
&&\hspace{1.35cm}+I_{16}(d)+I_{16}(a)+I_{17}(d)+I_{17}(a)\Big]
+\frac{f(f-1)(f-2)}{6}\Big[I_{18}(d)+I_{18}(a)\nonumber\\
&&\hspace{1.35cm}+I_{19}(d)+I_{19}+(a)I_{20}(d)+I_{20}(a)\Big]+\frac{f(f-1)(f-2)(f-3)}{24}\Big[I_{21}(d)
+I_{21}(a)\Big]\Big\}\Big|_{q=0}\,.\quad
\end{eqnarray}
Here, $I_i(d)$, $I_i(a)$ are integrals listed in the appendix.
Performing the integration according to~(\ref{rgdefin}) and evaluating the double $\varepsilon$, $\delta$-expansion, we find for the gyration radius:
\begin{eqnarray}
 \overline{\langle R^2_{g{\rm star}} \rangle}&=&\frac{dS}{6f}(3f-2)\left\{ 1+b\frac{2}{\varepsilon}-w\frac{2}{\delta}\right.\nonumber\\
 &&{}-\left.(b-w)\left[\frac{13}{12}+\frac{13}{2}\frac{(f-1)(f-2)}{3f-2}
 -\ln(2)\frac{4(f-1)(3f-5)}{3f-2}\right]\right\}.
\label{rstar}
\end{eqnarray}

The result for the radius of gyration of a single chain of total length $fS$ is straightforward:
 \begin{eqnarray}
 \overline{\langle R^2_{g{\rm chain}} \rangle} &=&\frac{d(Sf)}{6}\left[1+bf^{\varepsilon/2}\left( \frac{2}{\varepsilon}-\frac{13}{12}\right)
-wf^{\delta/2}\left( \frac{2}{\delta}-\frac{13}{12}\right) \right]\nonumber\\
&=&\frac{d(Sf)}{6}\left[1+b\left( \frac{2}{\varepsilon}-\frac{13}{12}\right)-w\left(
\frac{2}{\delta}-\frac{13}{12}\right)+(b-w)\ln(f) \right].
\label{rchain}
\end{eqnarray}

\begin{figure}[!t]
\begin{center}
\psfrag{s}{{\Large $s_1$}}
\psfrag{ss}{{\Large $s_2$}}
\includegraphics[width=13cm]{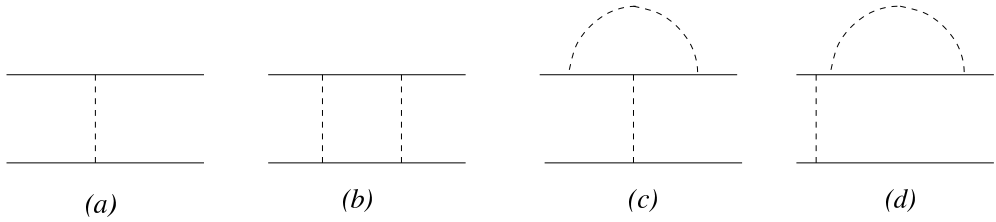}
 \end{center}
 \caption{Diagrammatic presentation of contributions into the partition function of two interacting chains.}
 \label{zll}
 \end{figure}

Finally, we need to perform renormalization of coupling constants.  To this end, we need the contributions to the partition function
$\overline{Z(S,S)}$ of a system of two interacting polymer chains of the same length $L$. In the diagrammatic representation of this
function, one takes only those terms into account
which contain at least one interaction line (see figure~\ref{zll}). In general, performing a thorough dimensional analysis of the
contributions produced by different diagrams, we find two
distinct classes of the diagrams. The first class of graphs
produces the terms which behave as $[S]^2[b_0]$, the sum of all such terms contributing to the function denoted by
$\overline{Z_{b_0}(S,S)}$; the second class of diagrams
behaves as
$[S]^2[w_0]$ and thus contributes to the function
$\overline{Z_{w_0}(S,S)}$.

We find the contributions to functions $\overline{Z_{b_0}(S,S)}$ and $\overline{Z_{w_0}(S,S)}$ as:
\begin{eqnarray}
\overline{Z_{b_0}(S,S)}&=&-b_0 S^2
\left[1-\frac{8b}{(2-d)(4-d)}+\frac{8w}{(2-a)(4-a)}\right.\nonumber \\
&&-32b\frac{d+2^{4-d/2}-10}{(d-2)(d-4)(d-6)(d-8)}\nonumber \\
&&-32\frac{w^2}{b}\frac{2a-d+2^{4-(2a-d)/2}-10}{(2a-d-2)(2a-d-4)(2a-d-6)(2a-d-8)}\nonumber \\
&&+\left.64w\frac{a+2^{4-a/2}-10}{(a-2)(a-4)(a-6)(a-8)}\right], \nonumber\\
\overline{Z_{w_0}(S,S)}&=&w_0 S^2
\left[1-\frac{8b}{(2-d)(4-d)}+\frac{8w}{(2-a)(4-a)}\right]. \label{zllw}
\end{eqnarray}
We may, therefore, define the dimensionless renormalized coupling constants $b_{\mathrm{R}}$, $w_{\mathrm{R}}$  by (\ref{ures}):
\begin{eqnarray}
&&b_{\mathrm{R}}=-\frac{ \overline{Z_{b}(S,S)}}{\langle\overline{Z(S)\rangle}^2}\left[2\pi\chi_0(b_0,w_0)S\right]^{-\frac{d}{2}},\nonumber\\
&&w_{\mathrm{R}}=-\frac{\overline{{Z}_{w}(S,S)}}{\langle\overline{Z(S)\rangle}^2}\left[2\pi\chi_0(b_0,w_0)S\right]^{-\frac{a}{2}}. \label{rebw}
\end{eqnarray}

The RG flows of the renormalized coupling constants are governed by corresponding $\beta$-functions:
\begin{eqnarray}
\beta_{b_{\mathrm{R}}}(b,w)=2S\frac{\partial  b_{\mathrm{R}}}{\partial S}\,,\qquad \beta_{w_{\mathrm{R}}}(b,w)=2S\frac{\partial w_{\mathrm{R}}}{\partial S}\,.
\label{bb}
\end{eqnarray}
Reexpressing in (\ref{bb}) $b$ and $w$ in terms of the renormalized couplings (\ref{rebw}), we
finally have:
\begin{eqnarray}
\beta_{b_{\mathrm{R}}}(b_{\mathrm{R}},w_{\mathrm{R}})&=&(4-d)b_{\mathrm{R}}-b_{\mathrm{R}}^2\frac{2d}{6-d}+w_{\mathrm{R}}b_{\mathrm{R}}\frac{2d}{6-a}+32b_{\mathrm{R}}^2\frac{d+2^{4-d/2}-10}{(d-2)(d-6)(d-8)}  \nonumber\\
&&+32w_{\mathrm{R}}^2\frac{2a-d+2^{4-(2a-d)/2}-10}{(2a-d-2)(2a-d-6)(2a-d-8)}
-64b_{\mathrm{R}}w_{\mathrm{R}}\frac{a+2^{4-a/2}-10}{(a-2)(a-6)(a-8)}\,, \nonumber\\
\beta_{w_{\mathrm{R}}}(b_{\mathrm{R}},w_{\mathrm{R}})&=&-(4-a)w_{\mathrm{R}}-w_{\mathrm{R}}^2\frac{2a}{6-a}+b_{\mathrm{R}}w_{\mathrm{R}}\frac{2a}{6-d}\,.\label{wwfinal}
\end{eqnarray}
Performing a double $\varepsilon$, $\delta$ expansion and keeping the terms up to linear in these parameters,
we then have the RG functions:
\begin{eqnarray}
&&\beta_b=\varepsilon b_{\mathrm{R}}-8b_{\mathrm{R}}^2-4w_{\mathrm{R}}^2+12b_{\mathrm{R}}w_{\mathrm{R}}\,,\nonumber\\
&&\beta_w=-\delta w_{\mathrm{R}}-4w_{\mathrm{R}}^2+4b_{\mathrm{R}}w_{\mathrm{R}}\,.\label{wwfinal1}
\end{eqnarray}
The fixed points $b_{\mathrm{R}}^*$, $w_{\mathrm{R}}^*$ of the renormalization group transformations are defined as the common zeros of
the RG functions (\ref{wwfinal1}). We find three distinct fixed points that determine the scaling behavior of the system in
 various regions of the $a$ and $d$ plane:
\begin{align}
&\text{Gaussian:} & &b_{\mathrm{R}}^*=0,\,w_{\mathrm{R}}^*=0 & &\text{stable at} \quad \varepsilon,\delta<0, \label{G}\\
&\text{Pure:} & &b_{\mathrm{R}}^*=\frac{\varepsilon}{8},\, w_{\mathrm{R}}^*=0& &\text{stable at} \quad \delta<\varepsilon/2,\phantom{55}\\
&\text{LR:} & &b_{\mathrm{R}}^*=\frac{\delta^2}{4(\varepsilon-\delta)},\,
w_{\mathrm{R}}^*=\frac{\delta(\varepsilon-2\delta)}{4(\delta-\varepsilon)} & &\text{stable at} \quad \varepsilon/2<\delta<\varepsilon.\phantom{55}
\label{R}
\end{align}
Here and below, the index ${\rm LR}$ means that the corresponding quantity is evaluated in the region of $d$, $a$ plane,
where the effect of long-range-correlated disorder is relevant.

We can also find the  estimates for critical exponents that govern the scaling of star polymers in solutions in the presence of a correlated disorder.
 Making use of definition (\ref{nuexp}), recalling the expression or renormalized scale (\ref{Lr}) and passing to the renormalized couplings, we have:
 \begin{equation}
 \nu(b_{\mathrm{R}},w_{\mathrm{R}})=\frac{1}{2}+\frac{b_{\mathrm{R}}}{2}-\frac{w_{\mathrm{R}}}{2}\, .
 \end{equation}
Evaluating this expression at fixed points of the renormalization group transformation (\ref{G})--(\ref{R}), we find the corresponding estimates
 for the size exponents:
 \begin{eqnarray}
&&\nu^{{\rm Gaussian}}=\frac{1}{2}\,,\label{nuG}\\
&&\nu^{{\rm Pure}}=\frac{1}{2}+\frac{\varepsilon}{16}\,,\\
&&\nu^{{\rm LR}}=\frac{1}{2}+\frac{\delta}{8}\,.\label{nuR}
\end{eqnarray}
Note that  these exponents govern the scaling behavior of macromolecules in the regions of stability of the corresponding fixed points (\ref{G})--(\ref{R}).

Similarly, evaluating (\ref{gammaexp}) and taking into account (\ref{Zfeps}), one finds in terms of renormalized couplings:
\begin{equation}
\gamma(b_{\mathrm{R}},w_{\mathrm{R}})=1+\frac{b_{\mathrm{R}}f(3-f)}{2}-\frac{w_{\mathrm{R}}f(3-f)}{2}\,.
\end{equation}
 Again, evaluating this expression at the fixed points (\ref{G})--(\ref{R}), we find:
 \begin{eqnarray}
&&\gamma_f^{{\rm Gaussian}}=1,\label{gammaG}\\
&&\gamma_f^{{\rm Pure}}=\frac{1}{2}+\frac{\varepsilon}{16}f(3-f),\\
&&\gamma_f^{{\rm LR}}=\frac{1}{2}+\frac{\delta}{8}f(3-f).\label{gammaR}
\end{eqnarray}
At $f=1$, one restores the corresponding exponents for a single polymer chain.
Note, that  the first-order expressions for fixed point coordinates (\ref{G})--(\ref{R}) and for critical exponents  (\ref{nuG})--(\ref{nuR}) and (\ref{gammaG})--(\ref{gammaR}) were obtained here for an annealed system. Thus, with $\nu^{{\rm LR}}$ and $\gamma_f^{{\rm LR}}$ as obtained in the regime of {\it annealed} disorder,
we restore the corresponding exponents that govern the scaling of polymers in solutions in the presence of a {\it quenched}
long-ranged correlated disorder studied by us earlier~\cite{Blavatska01a,Blavatska01a_2}. This supports the statement of equivalence
between the annealed and quenched averaging when dealing with polymer systems.

Finally, we obtain  estimates for the size ratio $g={\overline{\langle R^2_{g{\rm star}} \rangle}}\big/{ \overline{\langle R^2_{g{\rm chain}} \rangle}}$ of star and linear polymers, recalling (\ref{rstar}) and (\ref{rchain}):
\begin{equation}
  g(f)= \left\{
  \begin{array}{ll}
g^{{\rm Gaussian}},& \hbox{ $\varepsilon,\delta<0$},\\
g^{{\rm Pure}}, & \hbox{ $\delta<\varepsilon/2$},\\
g^{{\rm LR}},   & \hbox{ $\varepsilon/2<\delta<\varepsilon$}
  \end{array}\right.\label{}
  \end{equation}
with:
 \begin{eqnarray}
 g^{{\rm Gaussian}}&=&\frac{3f-2}{f^2}\,,\\
g^{{\rm Pure}}&=&\frac{3f-2}{f^2}\left\{1-\frac{\varepsilon}{8}
\left[\frac{13}{2}\frac{(f-1)(f-2)}{3f-2}- \ln(2)\frac{4(f-1)(3f-5)}{3f-2}+\ln(f)\right]\right\},\label{gratio_ex_pure}\\
g^{{\rm LR}}&=&\frac{3f-2}{f^2}\left\{1-\frac{\delta}{4} \left[\frac{13}{2}\frac{(f-1)(f-2)}{3f-2}- \ln(2)\frac{4(f-1)(3f-5)}{3f-2}+\ln(f)\right]\right\} \label{gratio_ex}.
\end{eqnarray}
With  $g^{{\rm Pure}}$, we restore the size ratio for the case of polymers in a pure solvent evaluated previously in references~\cite{Miyake82,Miyake82_2,Alessandrini92}.

\section{Conclusions and outlook}

In the present paper, we have studied the impact of structural disorder on the static scattering function of $f$-arm star branched polymers in $d$ dimensions.
To this end, we consider the model of a star polymer immersed in a good solvent  in the presence of structural defects,
correlated
at large distances according to (\ref{func}) with parameter $a$  ~\cite{Weinrib83}.
The impact of such long-range-correlated disorder on the scaling of linear and star branched polymers has been analyzed in
our  previous works~\cite{Blavatska01a,Blavatska01a_2}.
Here, we are interested, in particular, in the ratio $g(f)$
of scattering intensities of star and linear polymers of the same molecular weight, which is a universal experimentally measurable quantity
[see (\ref{exper})]. We used a direct polymer renormalization approach~\cite{desCloizeaux} and evaluated the results within the double $\varepsilon=4-d$, $\delta=4-a$-expansion.

\begin{figure}[!b]
\centerline{
\includegraphics[width=0.48\textwidth]{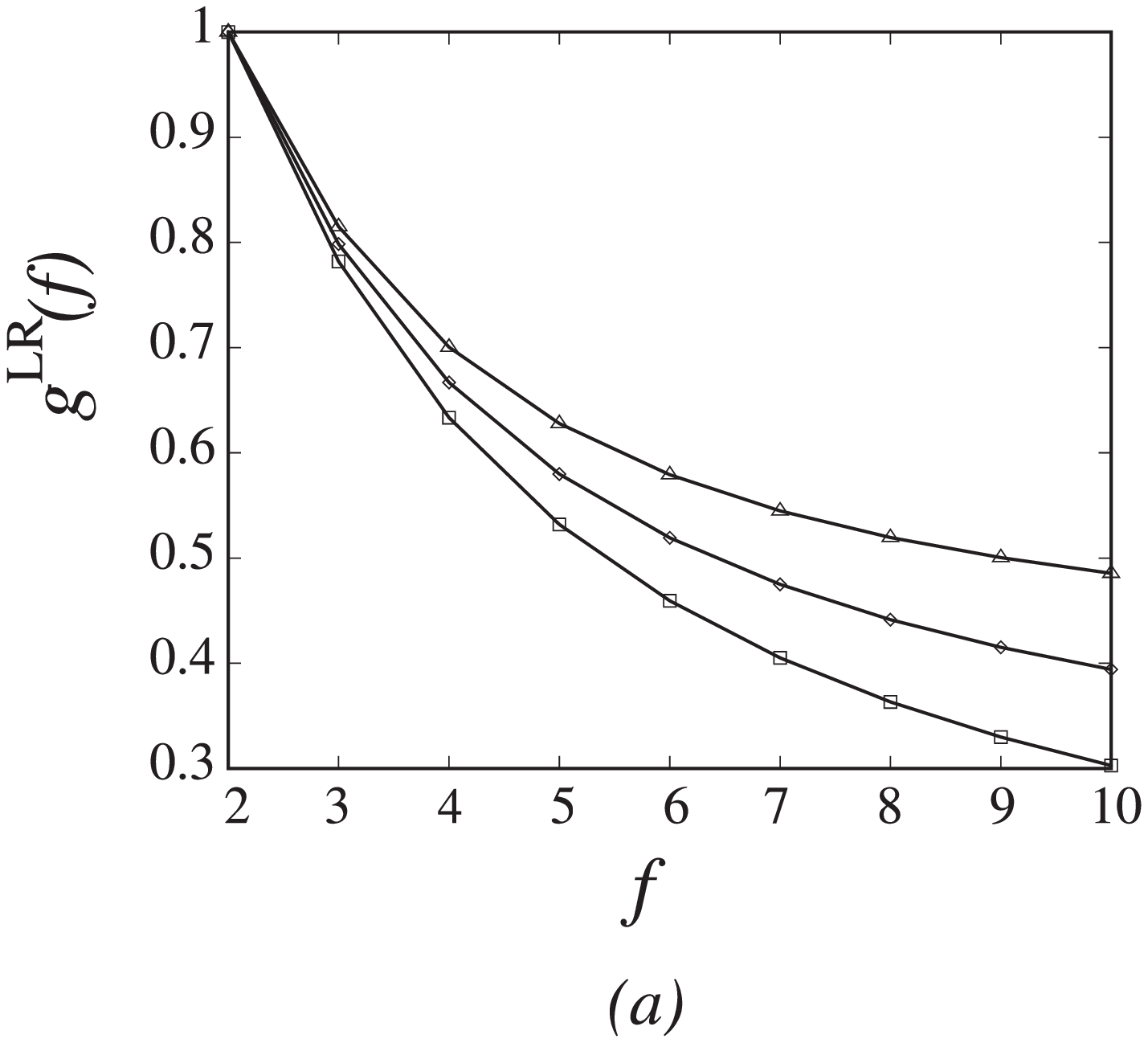}
 \hfill
\includegraphics[width=0.48\textwidth]{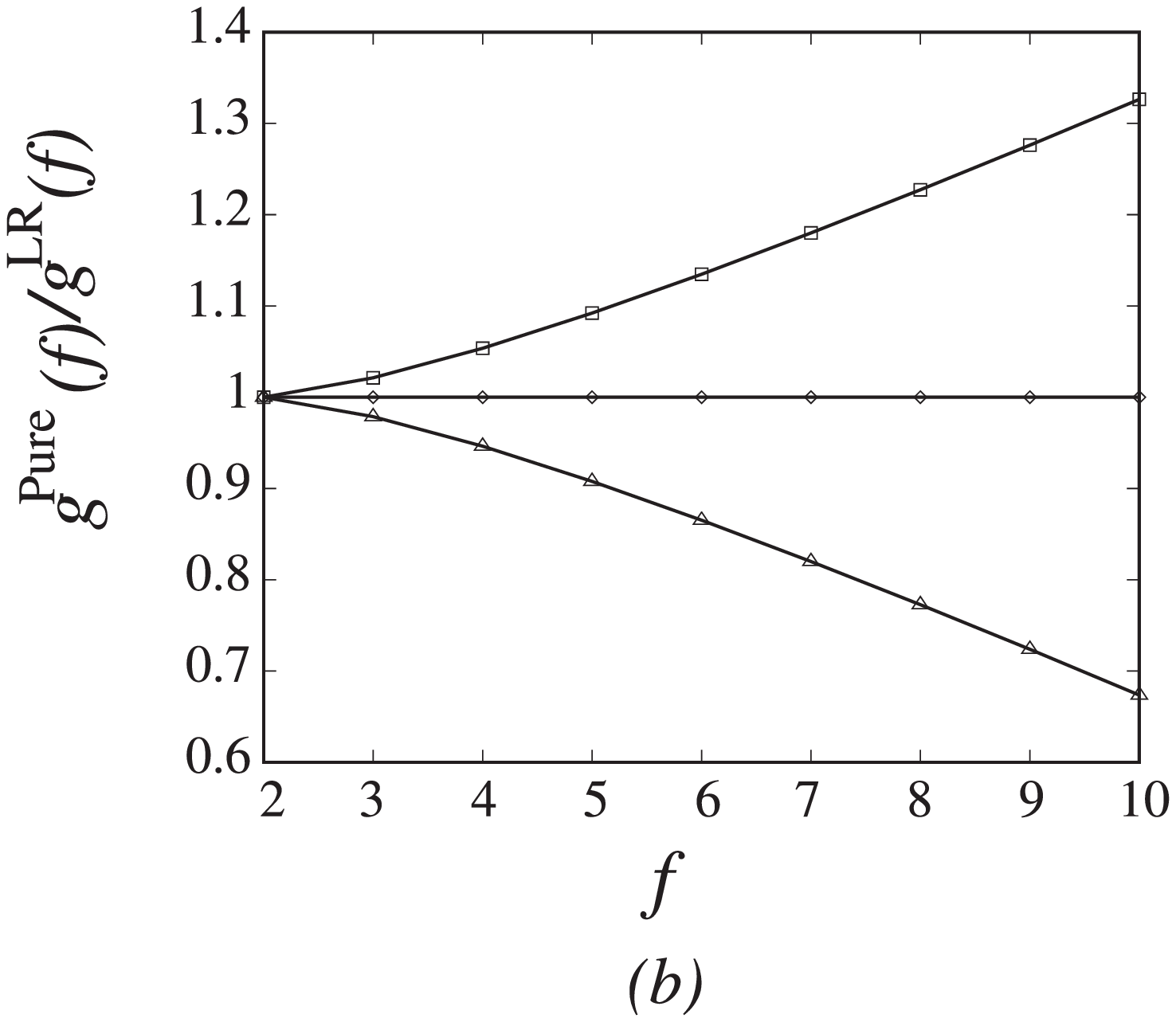}
}
 \caption{ Left: the size ratio (\ref{gratio_ex}) as a function of $f$ at different values of correlation parameter $\delta$. From below:  $\delta=0.1$, $\delta=0.5$, $\delta=1.0$. Right: the ratio~(\ref{gg_gg}) as a function of $f$ at different values of correlation parameter $\delta$. From above:  $\delta=0.1$, $\delta=0.5$, $\delta=1.0$. In both (a) and (b) we fix the value of parameter $\varepsilon=1$.}
\label{gg}
\end{figure}

Let us analyze the expressions obtained for the size ratio of star and linear polymers in a solution in the presence of structural defects.   First of all, as far as the $\delta=4-a$-prefactor in
(\ref{gratio_ex}) is positive for all $f>2$, one immediately concludes that the stronger the correlations of disorder (i.e., the larger is parameter $\delta$), the
larger  is the size ratio $g(f)$ and thus, the smaller is the distinction between
the size measure of a star and linear polymers of the same molecular weight.
In figure~\ref{gg}~(a), we present the estimates for the size ratio (\ref{gratio_ex})  as a function of the number of arms $f$ obtained by a direct
evaluation at several fixed values of the correlation parameter $\delta$. Besides an expected increase of $g^{{\rm LR}}(f)$ with growing
$\delta$ at each fixed $f$,
we also notice a decrease of the size ratio with $f$ for any fixed $\delta$. As can be seen from (4.40), this occurs already at the gaussian approximation: re-arranging the monomers of a linear chain into the shape of a star naturally leads to a smaller size and this effect becomes stronger when the star has more arms.

Another interesting aspect results from a qualitative comparison of the impact of a long-range-correlated disorder on the size ratio  with
the impact of the excluded volume effect in a pure solvent. We  consider the ratio:
\begin{equation}
\frac{g^{{\rm Pure}}(f)}{g^{{\rm LR}}(f)}=1-\left(\frac{\varepsilon}{8}-\frac{\delta}{4}\right)
\left[\frac{13}{2}\frac{(f-1)(f-2)}{3f-2}- \ln(2)\frac{4(f-1)(3f-5)}{3f-2}+\ln(f)\right]. \label{gg_gg}
\end{equation}
In figure~\ref{gg}~(b), we present the evaluation of this ratio as a function of $f$ at three
dimensions (taking $\varepsilon=1$) and several fixed values of correlation parameter $\delta$. Note, that a correlated disorder
with $\delta=0.5$ plays the role of ``marginal'', separating a region where the ratio
 (\ref{gg_gg}) increases with $f$ (at $\delta<0.5$) and a region where it decreases with $f$ (any $\delta>0.5$).

Let us recall that  with the critical exponents $\nu^{{\rm LR}}$ (\ref{nuR}) and $\gamma_f^{{\rm LR}}$ (\ref{gammaR}) that we
obtained in the present paper in the
regime of {\it annealed} disorder,
we restore the corresponding exponents that govern the scaling of polymers in solutions in the presence of a {\it quenched}
long-ranged correlated disorder, studied by us earlier~\cite{Blavatska01a,Blavatska01a_2}. This supports the statement of equivalence
between the annealed and quenched averaging when dealing with polymer systems. Thus, our qualitative estimates
for the size ratio  of star and linear polymers in annealed correlated disorder (\ref{gratio_ex}) also holds for the case of quenched systems.

Note that our results are based on the first-order perturbation theory expansions and provide a qualitative description of the impact of structural disorder on the quantities of interest. To obtain reliable quantitative estimates, the higher order analysis would be worthwhile, which is the subject of our forthcoming studies.

\section*{Acknowledgements}
This work was supported in part by the
FP7 EU IRSES project N269139  ``Dynamics and Cooperative Phenomena in Complex
Physical and Biological Media'' and Applied Research Fellowship of Coventry University.

\section*{Appendix}
\begin{figure}[b!]
\begin{center}
\psfrag{s}{{\Large$s'$}}
\psfrag{ss}{\hspace*{-0.2cm}{\Large $s''$}}
\psfrag{s1}{{\Large $s_1$}}
\psfrag{s2}{{\Large $s_2$}}
\includegraphics[width=3cm]{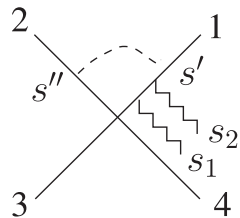}
  \end{center}
 \caption{Example of diagrammatic contribution into the radius of gyration of a star polymer (\ref{rgdefin}).}
 \label{di}
 \end{figure}

 Here, as an example we evaluate the analytic expression corresponding to diagram (9) in figure~\ref{Rstar} (shown  more in detail in figure~\ref{di})
 presenting contributions into the radius of gyration  of a star polymer. We have for this contribution:
 \begin{eqnarray} \nonumber
\overline {\langle {\rm e}^{-\ri\vec{q}[\vec{r}_1(s_2)-\vec{r}_1(s_1)]} \rangle}_{(9)}&=&-b_0\int{\cal D} r\, \exp\left\{ {- \frac{1}{2}\sum\limits_{i=1}^f\int\limits_0^{S}{\rm d}s \left[\frac{{\rm d} r_{i}(s)}{{\rm d} s}\right]^2}\right\}\\
&&{}\times{\rm e}^{-\ri\vec{q}[\vec{r}_1(s_2)-\vec{r}_1(s_1)]}
\int_{s_2}^{S}{\rm d}\,s{'}
\! \int_{0}^{S}\!{\rm d}\,s{''}\!\int {\rm d}\,{\vec k}\,
  {\rm e}^{-\ri\vec{k}[\vec{r}_1(s{'})-\vec{r}_2(s{''}) ]}\,.\nonumber
 \end{eqnarray}
 Rewriting  the last exponent:
 \begin{equation}
 -\ri\vec{k}\left[ \vec{r}_1(s{'})-\vec{r}_2(s{''}) \right]=-\ri\vec{k}\left\{ \left[\vec{r}_1(s{'})-\vec{r}_1(s_2)\right]+\left[\vec{r}_1(s_2)-\vec{r}_1(s_1)\right]
 +\left[\vec{r}_1(s_1)-\vec{r}_2(s{''})\right] \right\},
 \end{equation}
 and making use of (\ref{t1})--(\ref{t4}), one arrives at:
 \begin{eqnarray}
\overline {\langle {\rm e}^{-\ri\vec{q}[\vec{r}_1(s_2)-\vec{r}_1(s_1)]} \rangle}_{(9)}
=-b_0\int{\cal D} r\,
\exp\left\{ {- \frac{1}{2}\sum\limits_{i=1}^f\int\limits_0^{S}{\rm d}s \left[\frac{{\rm d} r_{i}(s)}{{\rm d} s}\right]^2}\right\}    I_9(d)\label{d9}
 \end{eqnarray}
 with:
 \begin{eqnarray}
I_9(d)&\equiv& \int\limits_{0}^{S}{\rm d}s''\int\limits_{s_2}^{S}{\rm d} s'\,\int {\rm d}\,{\vec k}\,
 {\rm e}^{-\frac{k^2}{2}\left(s^{'}-s_2+s_1+s^{''}\right)-\frac{(\vec{q}+\vec{k})^2}{2}(s_2-s_1)}\nonumber\\
&=&{\rm e}^{-\frac{q^2}{2}(s_2-s_1)}\int\limits_{0}^{S}{\rm d}s''\int\limits_{s_2}^{S}{\rm d} s'\,
\int {\rm d}\,{\vec k}\,{\rm e}^{-\frac{k^2}{2}\left(s^{'}+s^{''}\right)-\vec{q}\vec{k}(s_2-s_1)}\nonumber\\
&=&(2\pi)^{-d/2}{\rm e}^{-\frac{q^2}{2}(s_2-s_1)}\int\limits_{0}^{S}{\rm d}s''\int\limits_{s_2}^{S}{\rm d} s'\,
 \left(s''+s'\right)^{-d/2}{\rm e}^{\frac{q^2}{2}\frac{(s_2-s_1)^2}{\left(s'+s''\right)}}.
 \end{eqnarray}
 Taking the derivative over $q$ according to (\ref{def}) we get:
\begin{eqnarray}
\frac{\partial I_9(d)}{\partial q^2}\Bigg|_{q=0}&=&(2\pi)^{-d/2}d(s_2-s_1)\int\limits_{0}^{S}{\rm d}s''\int\limits_{s_2}^{S}{\rm d} s'\,
 \left(s''+s'\right)^{-d/2} \nonumber\\
 &&{}- (2\pi)^{-d/2}d(s_2-s_1)^2\int\limits_{0}^{S}{\rm d}s''\int\limits_{s_2}^{S}{\rm d} s'\,
 \left(s''+s'\right)^{-d/2-1}\nonumber\\
&=& (2\pi)^{-d/2}\frac{d(s_2-s_1)}{(1-d/2)(2-d/2)} \left[(2S)^{2-d/2} - (S+s_2)^{2-d/2}-S^{2-d/2}
+s_2^{2-d/2} \right]\nonumber\\
 &&{}-(2\pi)^{-d/2}\frac{d(s_2-s_1)^2}{(1-d/2)(-d/2)}\left [ (2S)^{1-d/2}-(S+s_2)^{1-d/2}-S^{1-d/2}+s_2^{1-d/2}\right].
  \end{eqnarray}
  Finally, the contribution into the gyration radius (\ref{rgdefin}) can be found by integrating over $s_1$, $s_2$.
Note, that the Gaussian integral in (\ref{d9}) cancels against the numerator
when averaging (\ref{aver}) is performed.

  The expressions corresponding to other diagrams in figure~\ref{Rstar} are listed below (note, that $I(d)$ and $I(a)$ arise, correspondingly,
  when treating interactions with couplings $b_0$ and $w_0$, respectively).
The factors $(2\pi)^{-d/2}$ and $(2\pi)^{-a/2}$ in front of each integral are omitted.
\begin{eqnarray*}
I_1(d)&=&{\rm e}^{-\frac{q^2}{2}(s_2-s_1)}\int\limits_{s_2}^{S}{\rm d}s''\int\limits_{s_1}^{s_2}{\rm d} s'\,
\left(s''-s'\right)^{-d/2}{\rm e}^{\frac{q^2}{2}\frac{(s_2-s')^2}{\left(s''-s'\right)}},\\
I_2(d)&=&{\rm e}^{-\frac{q^2}{2}(s_2-s_1)}\int\limits_{s_1}^{s_2}{\rm d}s''\int\limits_{0}^{s_1}{\rm d} s'\,
\left(s''-s'\right)^{-d/2}{\rm e}^{\frac{q^2}{2}\frac{(s''-s_1)^2}{\left(s''-s'\right)}},\\
I_3(d)&=&{\rm e}^{-\frac{q^2}{2}(s_2-s_1)}\int\limits_{s_2}^{S}{\rm d}s''\int\limits_{0}^{s_1}{\rm d} s'\,
\left(s''-s'\right)^{-d/2}{\rm e}^{\frac{q^2}{2}\frac{(s_2-s_1)^2}{\left(s''-s'\right)}},\\
I_4(d)&=&{\rm e}^{-\frac{q^2}{2}(s_2-s_1)}\int\limits_{s_2}^{S}{\rm d}s''\int\limits_{s_2}^{s''}{\rm d} s'\,
\left(s''-s'\right)^{-d/2},\\
I_5(d)&=&{\rm e}^{-\frac{q^2}{2}(s_2-s_1)}\int\limits_{0}^{s_1}{\rm d}s''\int\limits_{0}^{s''}{\rm d} s'\,
\left(s''-s'\right)^{-d/2},
\end{eqnarray*}
\begin{eqnarray*}
I_6(d)&=&{\rm e}^{-\frac{q^2}{2}(s_2-s_1)}\int\limits_{s_1}^{s_2}{\rm d}s''\int\limits_{s_1}^{s''}{\rm d} s'\,
 \left(s''-s'\right)^{-d/2}{\rm e}^{\frac{q^2}{2}(s''-s_1)},\\
 I_7(d)&=&{\rm e}^{-\frac{q^2}{2}(s_2-s_1)}\int\limits_{0}^{S}{\rm d}s''\int\limits_{0}^{s_1}{\rm d} s'\,
 \left(s''+s'\right)^{-d/2},\\
 I_8(d)&=&{\rm e}^{-\frac{q^2}{2}(s_2-s_1)}\int\limits_{0}^{S}{\rm d}s''\int\limits_{s_1}^{s_2}{\rm d} s'\,
 \left(s''+s'\right)^{-d/2}{\rm e}^{\frac{q^2}{2}\frac{(s'-s_1)^2}{(s'+s'')}},\\
 I_9(d)&=&{\rm e}^{-\frac{q^2}{2}(s_2-s_1)}\int\limits_{0}^{S}{\rm d}s''\int\limits_{s_2}^{S}{\rm d} s'\,
 \left(s''+s'\right)^{-d/2}{\rm e}^{\frac{q^2}{2}\frac{(s_2-s_1)^2}{(s'+s'')}},\\
I_{10}(d)&=&{\rm e}^{-\frac{q^2}{2}(s_2-s_1)}\int\limits_{0}^{S}{\rm d}s''\int\limits_{0}^{S}{\rm d} s'\, \left(s''+s'\right)^{-d/2},\\
 I_{11}(d)&=&{\rm e}^{-\frac{q^2}{2}(s_2-s_1)}\int\limits_{0}^{S}{\rm d}s''\int\limits_{0}^{s''}{\rm d} s'\,
 \left(s''-s'\right)^{-d/2},\\
 I_{12}(d)&=&{\rm e}^{-\frac{q^2}{2}(s_2+s_1)}\int\limits_{s_1}^{S}{\rm d}s''\int\limits_{s_1}^{s''}{\rm d} s'\,
 \left(s''-s'\right)^{-d/2},\\
 I_{13}(d)&=&{\rm e}^{-\frac{q^2}{2}(s_2+s_1)}\int\limits_{s_1}^{S}{\rm d}s''\int\limits_{0}^{s_1}{\rm d} s'\,
 \left(s''-s'\right)^{-d/2}{\rm e}^{\frac{q^2}{2}\frac{(s_1-s')^2}{\left(s''-s'\right)}},\\
 I_{14}(d)&=&{\rm e}^{-\frac{q^2}{2}(s_2+s_1)}\int\limits_{0}^{s_1}{\rm d}s''\int\limits_{0}^{s''}{\rm d} s'\,
 \left(s''-s'\right)^{-d/2}{\rm e}^{\frac{q^2}{2}\left(s''-s'\right)},\\
 I_{15}(d)&=&{\rm e}^{-\frac{q^2}{2}(s_2+s_1)}\int\limits_{s_2}^{S}{\rm d}s''\int\limits_{s_1}^{S}{\rm d} s'\,
 \left(s''+s'\right)^{-d/2}{\rm e}^{\frac{q^2}{2}\frac{(s_1+s_2)^2}{\left(s''+s'\right)}},\\
 I_{16}(d)&=&{\rm e}^{-\frac{q^2}{2}(s_2+s_1)}\int\limits_{0}^{s_2}{\rm d}s''\int\limits_{s_1}^{S}{\rm d} s'\,
 \left(s''+s'\right)^{-d/2}{\rm e}^{\frac{q^2}{2}\frac{(s_1+s'')^2}{\left(s''+s'\right)}},\\
 I_{17}(d)&=&{\rm e}^{-\frac{q^2}{2}(s_2+s_1)}\int\limits_{0}^{s_2}{\rm d}s''\int\limits_{0}^{s_1}{\rm d} s'\,
 \left(s''+s'\right)^{-d/2}{\rm e}^{\frac{q^2}{2}\left(s''+s'\right)},\\
 I_{18}(d)&=&{\rm e}^{-\frac{q^2}{2}(s_2+s_1)}\int\limits_{0}^{S}{\rm d}s''\int\limits_{0}^{s_2}{\rm d} s'\,
 \left(s''+s'\right)^{-d/2}{\rm e}^{\frac{q^2}{2}\frac{(s')^2}{\left(s''+s'\right)}},\\
 I_{19}(d)&=&{\rm e}^{-\frac{q^2}{2}(s_2+s_1)}\int\limits_{s_2}^{S}{\rm d}s''\int\limits_{0}^{S}{\rm d} s'\,
 \left(s''+s'\right)^{-d/2}{\rm e}^{\frac{q^2}{2}\frac{(s_2)^2}{\left(s''+s'\right)}},\\
 I_{20}(d)&=&{\rm e}^{-\frac{q^2}{2}(s_2+s_1)}\int\limits_{0}^{S}{\rm d}s''\int\limits_{0}^{s''}{\rm d} s'\,
 \left(s''-s'\right)^{-d/2},\\
%
 I_{21}(d)&=&{\rm e}^{-\frac{q^2}{2}(s_2+s_1)}\int\limits_{0}^{S}{\rm d}s''\int\limits_{0}^{S}{\rm d} s'\,
 \left(s''+s'\right)^{-d/2}.
\end{eqnarray*}

\newpage
\ukrainianpart

\title{Вплив безладу на статичну функцію розсіяння зіркових полімерів}
\author{В. Блавацька\refaddr{label1},  К. фон Фербер\refaddr{label2, label3},  Ю. Головач\refaddr{label1}}
\addresses{
\addr{label1} Інститут фізики конденсованих систем НАН України,  79011 Львів, Україна
\addr{label2} Дослідницький центр прикладної математики,  Університет Ковентрі,  CV1 5FB Ковентрі, Великобританія
\addr{label3} Інститут теоретичної фізики  II, Університет ім. Гайнріха Гайне,
D-40225 Дюссельдорф,  Німеччина}
%
%
%

\makeukrtitle

\begin{abstract}
\tolerance=3000%

Представлено аналіз впливу структурного безладу на
статичну функцію розсіяння $f$-гілкового зіркового полімера у $d$-вимірному просторі.
Розглянуто модель зіркового полімера  у хорошому розчиннику в присутності структурних дефектів,
скорельованих на великих віддалях  $r$ згідно степеневого закону  $ \sim r^{-a}$. Зокрема,
ми цікавимось відношенням $g(f)$
інтенсивностей розсіяння зіркового та лінійного полімерів однакової молекулярної маси, що є
універсальною, експериментально спостережуваною величиною. Ми застосовуємо метод прямого
полімерного перенормування і використовуємо подвійний $\varepsilon=4-d$, $\delta=4-a$-розклад.
Знайдено зростання величини $g(f)$ із зростанням параметра $\delta$. Таким чином, зростання кореляцій безладу
приводить до зменшення відмінності між розміром зіркових та лінійних полімерів із однаковою молекулярною вагою.

\keywords полімери, структурний безлад, універсальність, ренормалізаційна група
\end{abstract}
\end{document}